\documentclass[journal,transmag]{IEEEtran}

\usepackage{cite}
\usepackage{graphicx}
\usepackage{amssymb,amsmath}
\usepackage{xcolor}

\begin{document}

\title{Guidelines for interpreting microfocused Brillouin light scattering spectra}


\author{
	\IEEEauthorblockN{Nessrine Benaziz \IEEEauthorrefmark{1}, Thibaut Devolder\IEEEauthorrefmark{1}, Stéphane Andrieu\IEEEauthorrefmark{2}, Jamal Ben Youssef\IEEEauthorrefmark{3}  and Jean-Paul Adam\IEEEauthorrefmark{1}}
	\IEEEauthorblockA{\IEEEauthorrefmark{1} Centre de Nanosciences et de Nanotechnologies, CNRS, Université Paris-Saclay, Palaiseau, 91120,
    \\ France, nessrine.benaziz@universite-paris-saclay.fr}
	\IEEEauthorblockA{\IEEEauthorrefmark{2} Institut Jean Lamour, Université de Lorraine, Nancy, 54011, France}
    \IEEEauthorblockA{\IEEEauthorrefmark{3} LabSTICC, CNRS, Université de Bretagne Occidentale, 29285 Brest, France.}
}

\IEEEtitleabstractindextext{%
\begin{abstract}

We present an analysis of the influence of spin wave dispersion relations and profiles on microfocused Brillouin Light Scattering spectra. Three archetypal magnetic materials are reported: a 51-nm thick Bi-substituted YIG, a 25-nm thick Heusler compound and 50-nm thick CoFeB alloy. These samples were chosen because they exhibit strongly contrasting spectral features --peak frequencies, linewidth, skewness. The shapes of these spectral features reflect the underlying spin wave dispersion relations and the thickness profile of the related spin wave modes. While analytical expressions of the dispersion relations provide a satisfactory description of the spectra if the modes are in separate frequency domains, the exact dispersion relations and the exact mode profiles are required for a correct description of the spectra not only when mode hybridization is present in the range or near the range of frequencies and wavevector accessed by the experiment. Our examples of microfocused BLS spectra are handy references that can be used as interpretation guidelines for BLS spectra recorded on a broader range of materials.

\end{abstract}

}

\maketitle

\pagestyle{empty}
\thispagestyle{empty}

\IEEEpeerreviewmaketitle

\section{Introduction}

\IEEEPARstart{B}{rillouin} light scattering (BLS) is a renowned technique for probing spin waves (SW), the elementary magnetic excitations, and their associated quasiparticles, magnons, through their inelastic interaction with photons. BLS is commonly used in the wavevector-resolved mode (k-BLS), where a single well-defined SW wavevector is probed. In the microfocused BLS ($\mu$-BLS) configuration, the beam is focused through a high numerical aperture (NA) lens, which allows a wide range of in-plane SW wavevectors to be simultaneously probed within a single acquisition. The resulting spectra are often complex and require modeling to disentangle the different physical contributions. To facilitate the interpretation of these spectra, we developed a versatile analysis method \cite{benaziz_method_2025}. Here we illustrate our approach on three magnetic systems chosen because they exhibit strongly contrasting spectral features –peak frequencies, linewidth, skewness. The analysis of these three cases provides clear guidelines for the interpretation of BLS spectra recorded on a broader range of materials.

\section{Model accounting for the scattering spectrum of a population of spin waves}
We first briefly summarize the model of ref.~\cite{benaziz_method_2025} used to calculate $\mu$-BLS spectra. We consider a linearly polarized Gaussian beam focused through a high numerical aperture (NA) objective onto an in-plane magnetized film surface. This produces a cone of light with aperture $\theta_\text{max} = \arcsin(\text{NA})$. This angle defines the range of wavevectors involved in the $\mu$-BLS process. 
The microfocused spectra are approximated as the sum of the photon sole back-scattering events. The $\mu$-BLS spectra are then modeled by including the magnon population, the instrumental response function, the scattering intensity, and the spectral broadening. The input of the model is a full description of the spin wave dispersion relations and their profile across the thickness of the magnetic film. This contribution can be calculated using a numerical solver or an analytic formalism such as the Kalinikos-Slavin formalism \cite{kalinikos_theory_1986} under some restrictions that will be discussed in the last section. The model also accounts for the optical properties of the sample throughout the permittivity tensor. Quadratic magneto-optical Kerr effects are neglected, and multiple reflections of the light within the material are not taken into account.

\section{Magnetic and optical properties of the samples} \label{samples}
Three magnetic films were investigated. 

The first is an insulating bismuth-substituted yttrium iron garnet (BiYIG) film with composition $\text{Y}_{2.5}\text{Bi}_{0.5}\text{Fe}_5\text{O}_{12}$, grown on a (111)-oriented gadolinium gallium garnet substrate by Liquid Phase Epitaxy. 
The reported $\mu$-BLS spectra will be calculated assuming that its magnetic properties are a saturation magnetization $M_s = 200~\mathrm{kA/m}$, an exchange stiffness $A_{\mathrm{ex}} = 4~\mathrm{pJ/m}$, and a Landé factor $g = 2.08$. As for the other samples, these magnetic properties were inferred from various magnetometry experiments, including ferromagnetic resonance and angle-resolved hysteresis loops. 
The off-diagonal permittivity was taken to be -0.02 - 0.02i from ref.~\cite{wittekoek_magneto-optic_1975-}. The diagonal permittivity will be fitted from the respective sizes of the peaks in the forthcoming $\mu$-BLS spectra. We will converge to a value of 6.21+i.

The second system is a half-metallic 20-nm-thick Heusler compound of composition $\text{Co}_2\text{Mn}\text{Al}$  film grown by Molecular Beam Epitaxy on a MgO(001) substrate with a 10-nm MgO buffer layer and capped with MgO(3 nm)/Ti(2 nm)/Au(3 nm). Its magnetic parameters are assumed to be $M_s = 889~\mathrm{kA/m}$, cubic anisotropy field $H_K = 2~\mathrm{kA/m}$, $A_{\mathrm{ex}} = 7.8~\mathrm{pJ/m}$, and $g = 2.08$. We measured a diagonal permittivity of -3.49 + 6.60i by ellipsometry at 532 nm. The off-diagonal permittivity of 0.435 - 0.107i was taken from the literature at this same wavelength\cite{fu_dielectric_1995}. 

The third system is a metallic 50-nm-thick $\text{Co}_{40}\text{Fe}_{40}\text{B}_{20}$ (CoFeB) film with a nominal structure of Ta (3nm)/CoFeB (50 nm)/Ta (3nm). Its magnetic parameters are 
$M_s = 1.41 \times 10^6~\mathrm{A/m}$, $A_{\mathrm{ex}} = 16~\mathrm{pJ/m}$, and $g = 2.08$. The diagonal permittivity was measured by ellipsometry at 532 nm to be -5.41 + 13.69i. The off-diagonal permittivity is 0.435-0.107i and was taken from ref.~\cite{fu_dielectric_1995}.

\section{Experimental BLS spectra}
\begin{figure*}
    \centering
    \includegraphics[scale=0.1,width =17cm]{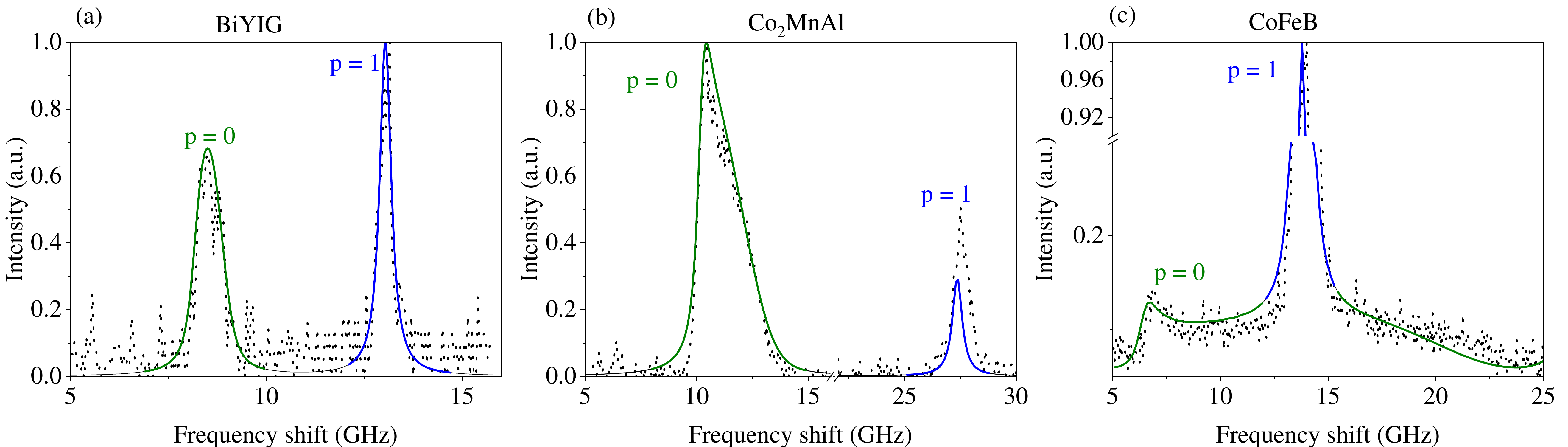}
    \caption{Experimental (black dots) and calculated (solid lines) $\mu$-BLS spectra measured on (a): a 51-nm-thick BiYIG film, (b): a 20-nm-thick $\text{Co}_2\text{Mn}\text{Al}$ film, and (c): a 50-nm-thick CoFeB film. The $p = 0$ peak is highlighted in green and the $p=1$ in blue.}
    \label{Imag1}
\end{figure*}
Figure~\ref{Imag1} collects the $\mu-$BLS spectra of the three samples at 300 K. 
The BiYIG sample [Fig.~\ref{Imag1}(a)] was measured at an applied field of 200 mT and with a microscope of NA=0.75. The $\text{Co}_2\text{Mn}\text{Al}$ [Fig.~\ref{Imag1}(b)] was measured under 100 mT with a slightly lower NA of 0.55. The CoFeB sample [Fig.~\ref{Imag1}(c)] was measured in 30 mT using a NA of 0.75. 

The frequency windows are chosen to include the first two peaks of each spectrum. 
The lowest frequency peak (green in Fig.~\ref{Imag1}, labeled $p = 0$) corresponds to the SW branch that emerges from the ferromagnetic resonance mode and has the most uniform mode profile across the film thickness. The higher frequency peak (blue in Fig.~\ref{Imag1}, labeled $p = 1$) corresponds to the branch with a perpendicular standing spin-wave (PSSW) character.
The $p = 1$ peak appears symmetric for all three samples. On the other hand, the $p = 0$ peak shows markedly different lineshapes.

For the BiYIG film the $p = 0$ peak is symmetric and narrow. In contrast, the $\text{Co}_2\text{Mn}\text{Al}$ film exhibits a broad $p = 0$ peak that is skewed towards high frequencies. Finally in the $\text{Co}_{40}\text{Fe}_{40}\text{B}_{20}$ film, the $p = 0$ peak is very broad and spans from 6 to 25 GHz. It appears as a shallow asymmetric hill with a long high-frequency tail.

The spectra calculated using the model of ref.~\cite{benaziz_method_2025} and the material properties listed in section \ref{samples} are compared with the experimental results in Fig. \ref{Imag1}. The model is in satisfactory agreement with the experiment. In particular, it reproduces the spectral peak shapes and their relative intensities. The difference in amplitude may arise from errors in the optical constants or from effects that are neglected in our model, such as the photon scattering in non-back-scattering directions or the multiple reflections of the photons within the samples.

\section{Understanding the shapes of the peaks in microfocused BLS spectra}
\begin{figure*}
    \centering
    \includegraphics[width =17cm]{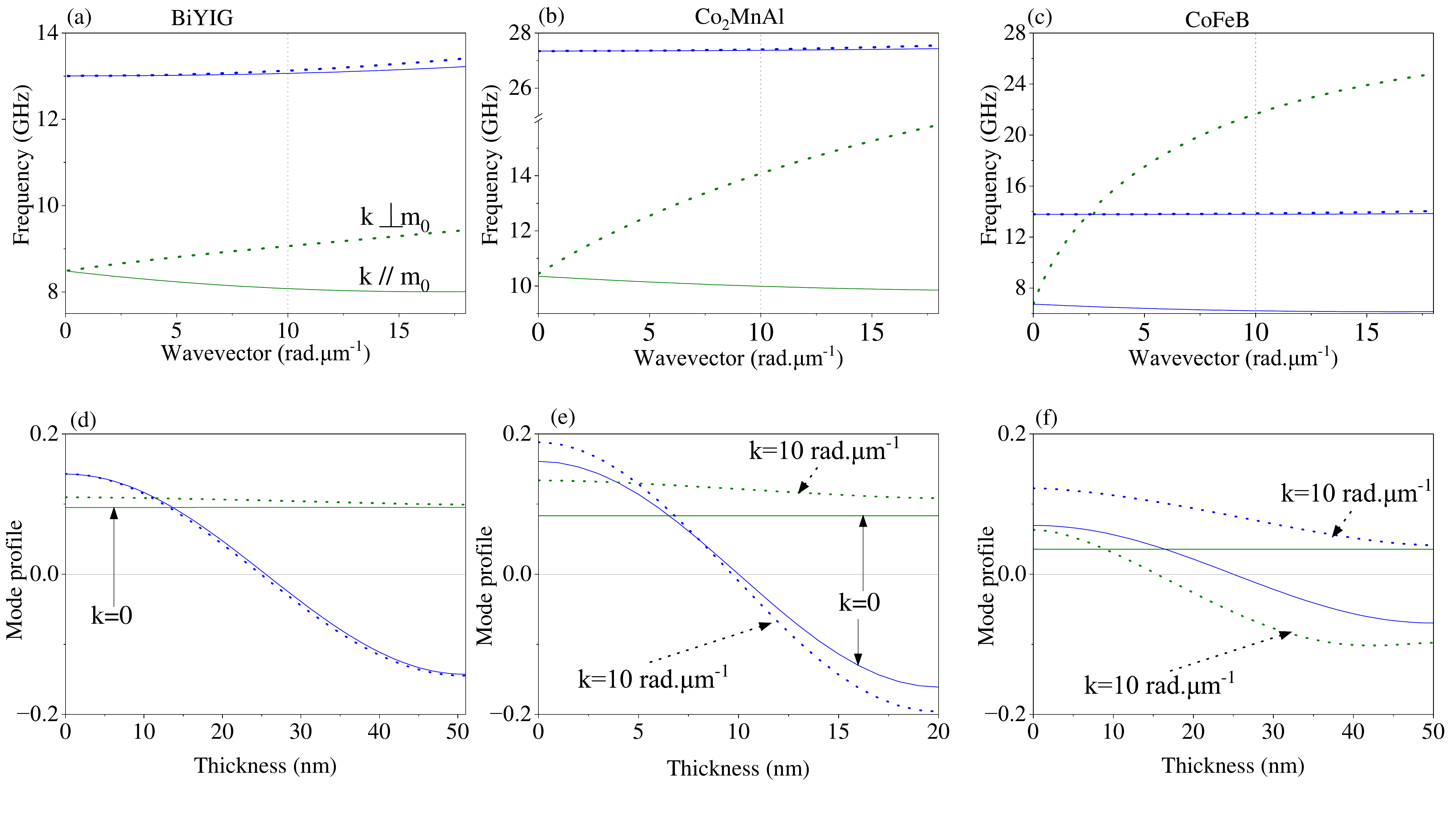}
    \caption{ TetraX simulations of spin wave properties. Top row: Dispersion relations in Damon-Eshbach geometry (dotted lines) and backward volume (solid lines) geometry for the (a) BiYIG (b) $\text{Co}_2\text{Mn}\text{Al}$ and (c) CoFeB sample. Bottom row: Numerically obtained thickness profiles of the spin waves at a $k = 0 ~\text{rad}/\mu\text{m}$ (solid lines) and at $k = 10 ~\text{rad}/\mu\text{m}$ (dashed lines) for the (d) BiYIG (e) $\text{Co}_2\text{Mn}\text{Al}$ (f) CoFeB sample. The green curves indicate the modes contributing to the $p=0$ peak, while the blue curves indicate the modes contributing to the $p = 1$ peak. The $m_0$ is for static magnetization.  }
    \label{Imag2}
\end{figure*}

As mentioned in \cite{benaziz_method_2025}, the spin wave density of states (DOS) and the $\mu$-BLS spectra are related. Since the DOS reflects the dispersion relation, the latter is systematically useful to understand a $\mu$-BLS spectrum.
Fig.~\ref{Imag2} shows the dispersion relations of the backward-volume magnetostatic spin waves (BVMSW), magnetostatic surface spin waves (MSSW), and the first PSSW mode, calculated numerically using the TetraX eigenmode solver \cite{korber_finite-element_2021,korber_numerical_2021} for (a) BiYIG, (b) $\text{Co}_2\text{Mn}\text{Al}$ and (c) CoFeB films.

The BiYIG sample exhibits almost flat dispersion relations which result in a sharply peaked DOS. This concentration of states at two single frequencies enhances the scattering intensity, resulting in two narrow and well-defined $\mu$-BLS peaks.

Similarly, the dispersion relation of the BVMSW mode of $\text{Co}_2\text{Mn}\text{Al}$ is also nearly flat, but its MSSW counterpart has a large group velocity, because of the large magnetization of this compound. As a result, the MSSW contribution to the DOS spreads over a broad range of frequencies. This broad range of frequencies accounts for the observed peak broadening and peak asymmetry in the $\mu$-BLS spectrum of Co$_2$MnAl [Fig.~\ref{Imag1}(b)].

A notable feature is observed in the spectrum of the CoFeB film. This arises because the MSSW and the first PSSW mode anticross at a wavevector of $k = 2.5~\text{rad}/\mu\text{m}$, a value that lies \textit{within} the range of wavevector relevant to the $\mu$-BLS spectra. This anticrossing produces the composite shape in the $\mu$-BLS spectrum, with the overlapping of the $p=0$ and $p=1$ peaks [Fig.~\ref{Imag1}(b)]. \\
Besides, this hybridization mixes the two modes and modifies their respective thickness profiles. The mode profiles of the FMR and the first PSSW mode at $k = 0$, and of the MSSW and first PSSW mode at $k = 10~\text{rad}/\mu\text{m}$ in the Damon Eshbash (DE) geometry are shown in Fig.~\ref{Imag2}(d--f) for the BiYIG, $\text{Co}_2\text{Mn}\text{Al}$, and CoFeB films, respectively. \\
As expected, at $k = 0$ the three samples exhibit very same mode profiles with the FMR mode having a uniform amplitude across the thickness, and the first PSSW mode displaying a cosinusoidal profile with a node at the center of the film. 

At  $k = 10 ~\text{rad}/\mu\text{m}$, the MSSW modes from the BiYIG and $\text{Co}_2\text{Mn}\text{Al}$ films show a profile that is more localized near the surface. Yet, as noted by \cite{kostylev_non-reciprocity_2013} the profile is distinct from the exponential decay predicted in the Damon-Eshbach model \cite{damon_magnetostatic_1961-1}. The first PSSW modes keep a cosine profile.
The mode profiles of the CoFeB sample exhibit a distinct behavior. After the hybridization, the modes acquire a mixed character. The branch that should correspond to the MSSW mode displays a profile with an off-centered node and the mode associated with the first PSSW exhibits a surface-localized profile.

\section{Effect of the film thickness on the peak shapes}

\begin{figure*}
    \centering
    \includegraphics[width =17cm]{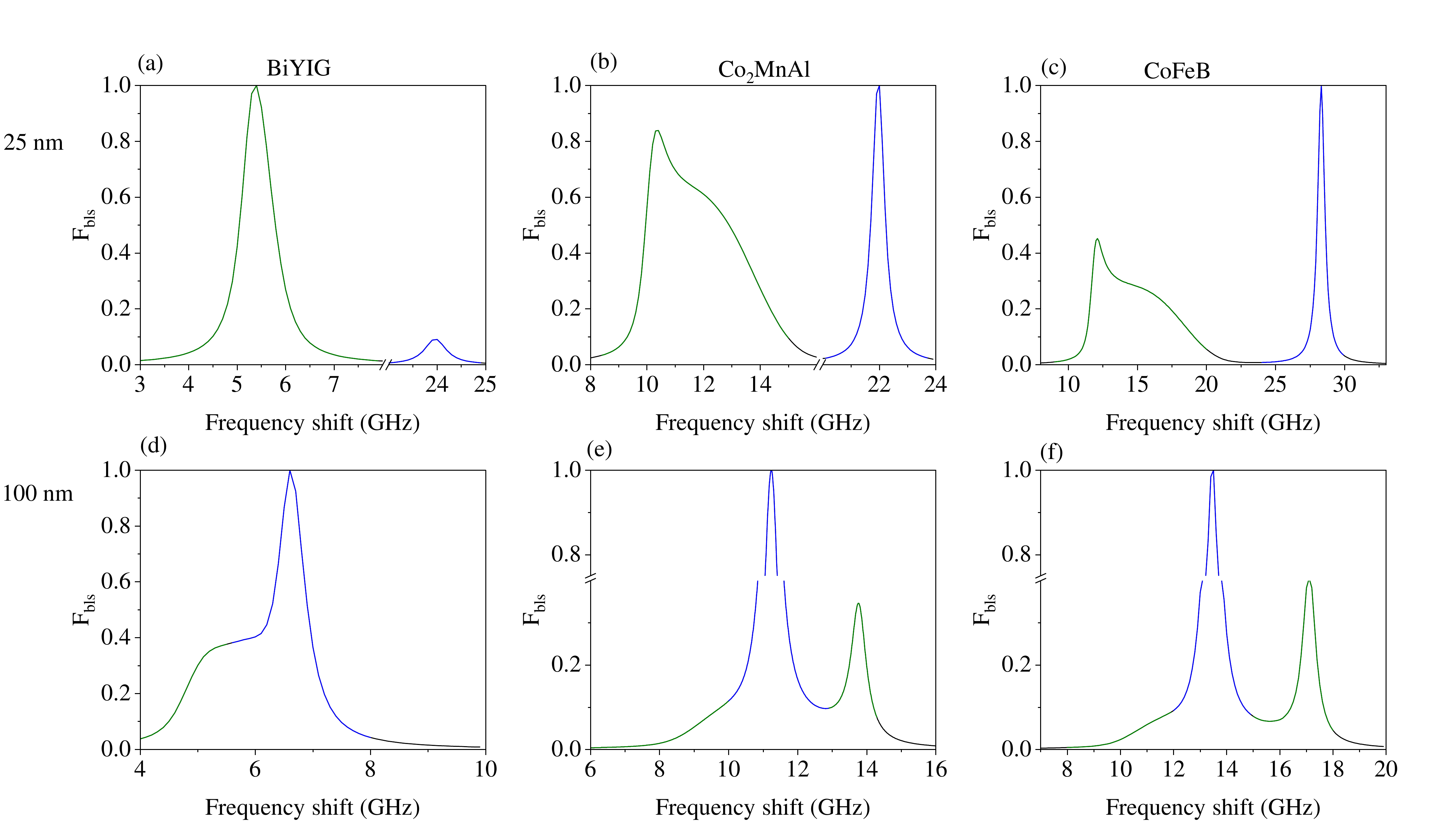}
    \caption{Comparative influence of film thickness on the shapes of $\mu$-BLS anti-Stokes spectra. Top row: for 25-nm-thick films of (a) BiYIG, (b) $\text{Co}_2\text{Mn}\text{Al}$  and (c) CoFeB. Bottom row: for 100-nm-thick films of (d) BiYIG, (e) $\text{Co}_2\text{Mn}\text{Al}$  and (f) CoFeB.}
    \label{Imag4}
\end{figure*}

The $\mu$-BLS spectra exhibit a strong dependence on the dispersion relation, which in turn depends on the group velocity, $v_g = \frac{\partial \omega}{\partial k}$. In this section, we will concentrate on the non-trivial dependence of $\mu$-BLS spectra on the sample thickness, as the group velocity can vary significantly due to changes in the sample's thickness.
Figure~\ref{Imag4} compares the spectra for 25-nm and 100-nm-thick films expected for a numerical aperture of 0.75 and an applied field of 100 mT. The spectra obtained for these two thicknesses are markedly different even when they correspond to the same material.

For the 25-nm-thick films [Figs.~\ref{Imag4}(a–c)], the $\mu$-BLS spectra systematically exhibit well-separated peaks corresponding to distinct SW branches. A shoulder is present in the low frequency peaks when the Damon-Eshbach spin waves have a large group velocity. In contrast, the 100-nm-thick films [Figs.~\ref{Imag4}(d–f)] systematically display overlapping peaks.

\begin{figure}
    \centering
    \includegraphics[width =9cm]{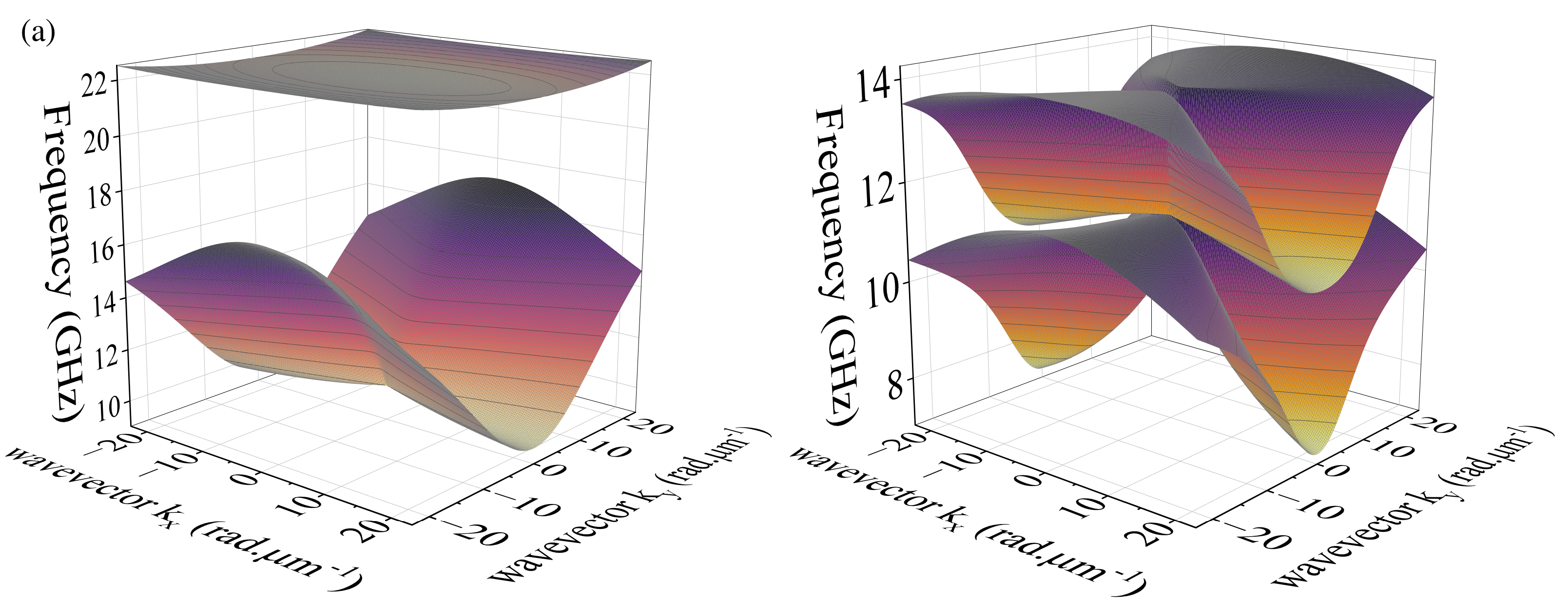}
    \caption{ Spin wave dispersion relation for all propagation directions for in-plane magnetized 25-nm-thick Heusler film (a) and 100-nm-thick Heusler film. }
    \label{Imag5}
\end{figure}
This transition from separated peaks to overlapping peaks is further illustrated in Fig.~\ref{Imag5}, which presents the spin wave dispersion relations for \textit{all} in-plane propagation directions, as calculated from TetraX. For the 25-nm-thick $\text{Co}_2\text{Mn}\text{Al}$ layer (Fig.~\ref{Imag5}(a)), two distinct SW dispersion surfaces are observed. The surface at a higher frequency presents a low $v_g$, while the lower-frequency surface exhibits a pronounced curvature with a saddle-like shape. For the 100-nm-thick $\text{Co}_2\text{Mn}\text{Al}$ layer [Fig.~\ref{Imag5}(b)], the two dispersion surfaces get closer together and would intersect if there was no dipolar coupling between the unperturbed modes. An avoided crossing happens and the two spin wave surfaces are now both strongly curved and have some overlap in their frequency span. 

\begin{figure}
    \centering
    \includegraphics[width =9cm]{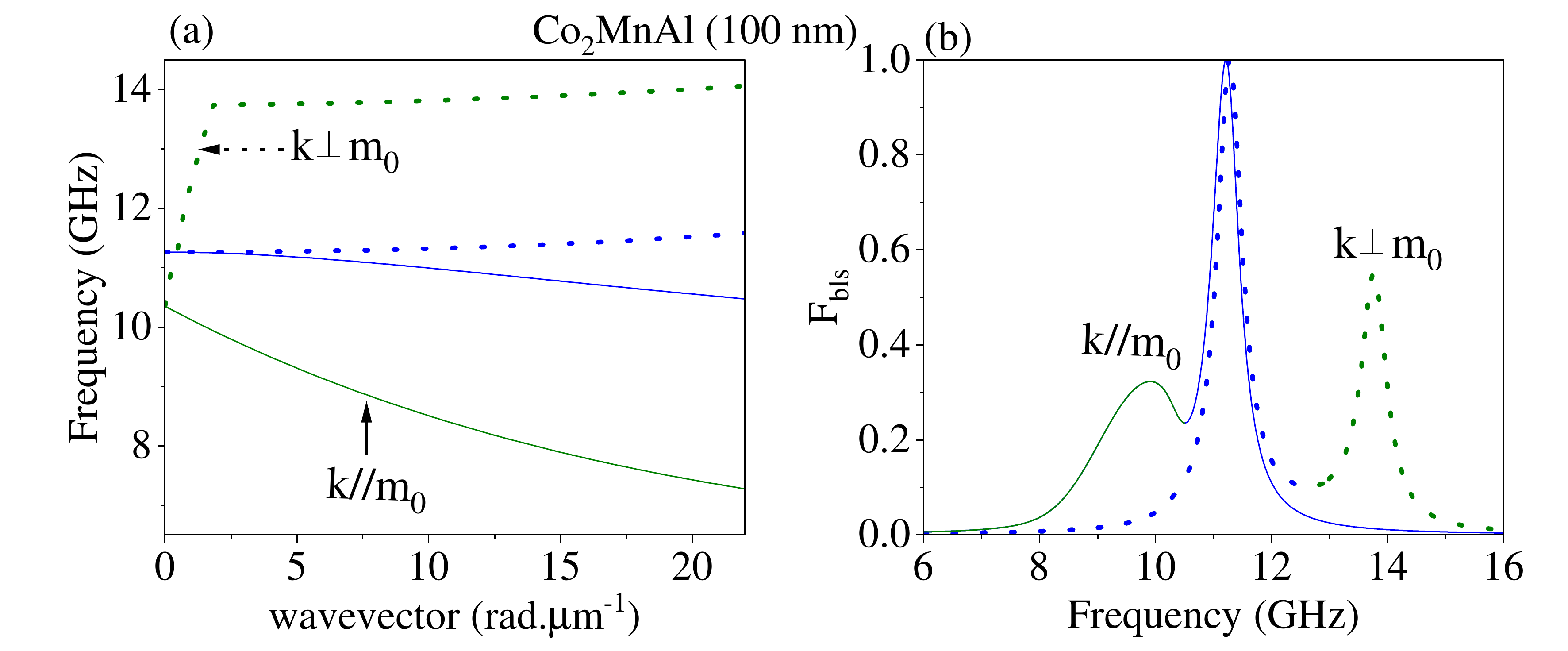}
    \caption{Comparative contributions of the spin waves of selected wavevector orientations. (a): Dispersion relation in the Damon-Eshbach (dot line) and Backward Volume (bold line) geometry. (b): Partial $\mu$-BLS anti-Stokes spectra obtained if the only spin waves populated in the material would be the backward waves (solid line) and the Damon-Eshbach (dashed line). The contributions to the $p = 0$ peak are in green while in blue for the $p = 1$ peak.}
    \label{Imag6}
\end{figure}

For the 100-nm-thick $\text{Co}_2\text{Mn}\text{Al}$ and CoFeB films [Figs.~\ref{Imag4}(e–f)], two distinct peaks are observed in the $\mu$-BLS spectra. In contrast to the naive --but usually true-- expectation that the lowest frequency peak would originate from the ferromagnetic resonance mode, here the lowest frequency peak is roughly at the frequency of the $ k = 0 $ PSSW mode, while the high frequency peak corresponds to the MSSW mode at $k = 1.84$ rad.$\mu$$\text{m}^{-1}$, where the group velocity vanishes and the dispersion relation changes from a positive slope to flat [Fig.~\ref{Imag6}(a)]. 
To support this interpretation, we computed partial $\mu$-BLS spectra for SW propagating in specific directions. For SWs propagating perpendicular to the static magnetization, the spectra exhibit two peaks [Fig.~\ref{Imag6}(b)]: a low-frequency peak at the PSSW 1 frequency and a high-frequency peak at the $k = 1.84$ rad.$\mu$$\text{m}^{-1}$ MSSW frequency. For waves parallel to the magnetization, the PSSW 1 peak is observed along with a weak low-frequency feature from the negative-$v_g$ BVSW mode [Fig.~\ref{Imag6}(b)]. 
We emphasize that the calculations for the 100-nm film thickness shown in Figures~\ref{Imag5} and \ref{Imag6} are partial calculations intended to illustrate an example and do not reflect the full spectrum as only the two lowest frequency modes were considered. The seemingly abrupt change in the dispersion at $k = 1.84$ rad.$\mu$$\text{m}^{-1}$ [Fig~\ref{Imag6}(a)] arises from the anti-crossing between the MSSW and the (not shown) second PSSW 2 which falls within the wavevector range of the experiment for this film thickness.

\section{Comparison of analytical vs numerical input}

So far we have been using mode profiles and mode frequencies numerically evaluated using TetraX, which is numerically heavy. It is interesting to study whether an analytical option can be used. In this section, we discuss whether the Kalinikos-Slavin (KS) model \cite{kalinikos_theory_1986} can be used. We focus on the 25-nm-thick films since at this thickness no mode crossing occurs [Fig. \ref{Imag6}] and the eigenfrequencies of the KS mode are generally thought to be accurate.

\begin{figure}
    \centering
    \includegraphics[width =9cm]{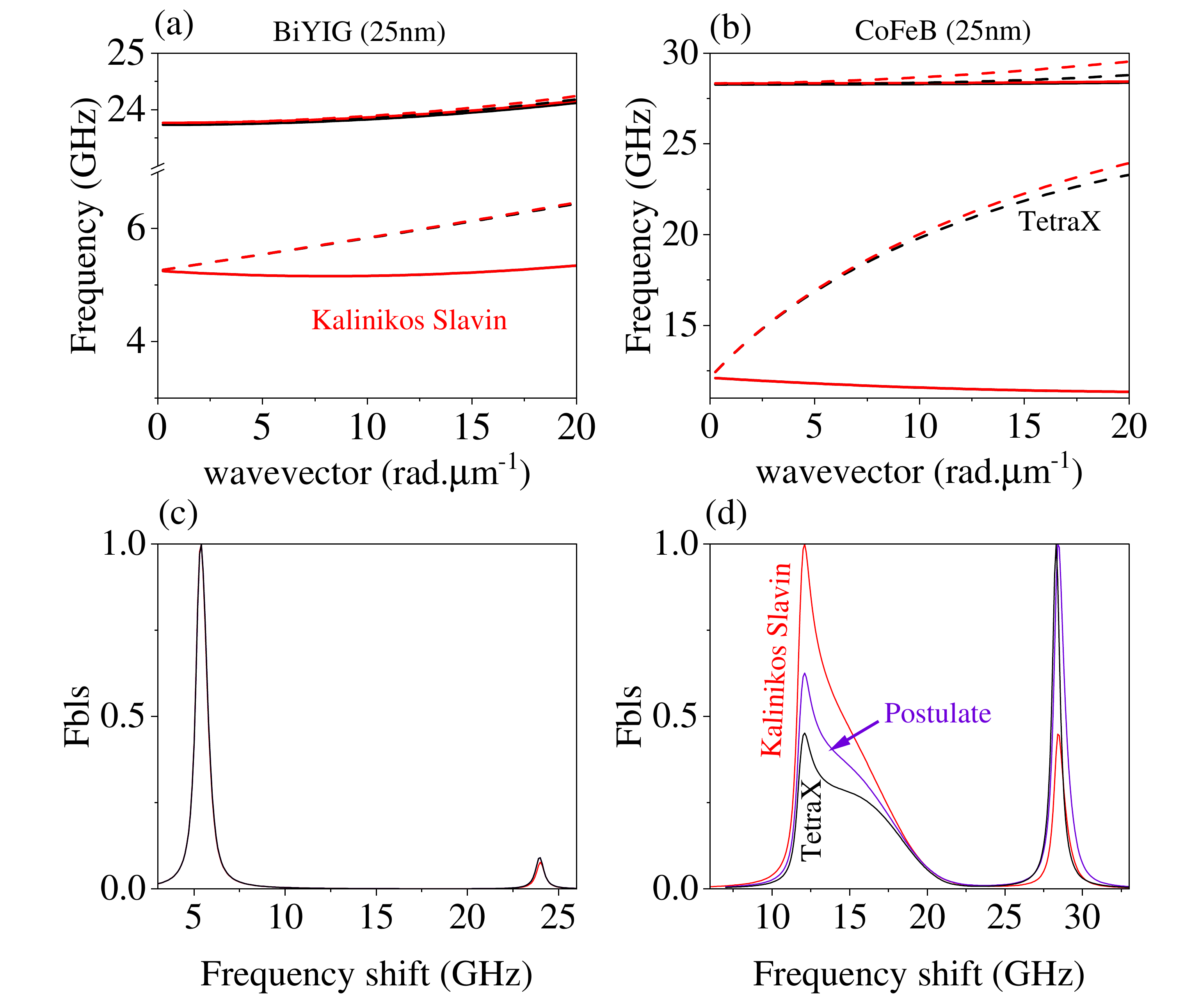}
    \caption{Comparison between the dispersion relations and the corresponding $\mu$-BLS spectra calculated with spin wave properties either from the Kalinikos-Slavin framework (Eq. 45 and Eq 46 of ref.  \cite{kalinikos_theory_1986}, red curves) and from the numerical eigensolver TetraX \cite{korber_finite-element_2021,korber_numerical_2021} (black curves).
(a,b): Dispersion relations with solid lines corresponding to backward-volume geometry and dashed lines corresponding to Damon–Eshbach geometry. 
(c,d) Corresponding $\mu$-BLS spectra. The purple curve is for mode profiles according to Eq.~\ref{postulat} and mode frequencies according to the KS theory.}
    \label{Imag7}
\end{figure}

Figure~\ref{Imag7} compares the dispersion relations calculated using TetraX with those obtained from the zeroth order expressions [Eqs.~(45)-(46)] of the KS model. For the BiYIG [Fig.~\ref{Imag7}(a)], the frequencies from TetraX and KS essentially match. The agreement is still satisfactory for the CoFeB film [Fig.~\ref{Imag7}(b)], but the frequencies calculated using TetraX are slightly lower than those predicted by the KS model.

The mode profiles are required to compute the $\mu$-BLS spectra. In \cite{kalinikos_theory_1986}, the mode profiles in the zero-order approximation are given by Eqs.~(13),(14),(20) and (21). In these expressions, the precession is circular (the ellipticity $\epsilon$ of the modes is one) and the low-frequency mode is assumed uniform across the thickness regardless of the SW propagation direction. The corresponding $\mu$-BLS spectra are compared with those obtained using TetraX input in Fig.~\ref{Imag7}. A good agreement is observed for the BiYIG sample [Fig.~\ref{Imag7}(c)], whereas the KS explicit expressions completely fail to account for the spectrum of the CoFeB sample [Fig.~\ref{Imag7}(d)]. 

One may hypothesize that this failure is due to the assumption of circular precession and of a uniform thickness profile of the low frequency modes. To test this hypothesis, we attempted to recompute the $\mu$-BLS spectra by keeping the KS frequencies but modifying the modes in an ad-hoc manner by including an ellipticity factor and postulating an exponential mode profile for the low-frequency dispersion surface, as in the Damon-Eshbach theory \cite{damon_magnetostatic_1961-1}.

The film lies in the $(xy)$ plane with the static magnetization being along the $x$ axis and $z$ the normal axis. The wavevector $\vec k$ forms an angle $\phi$  with the $x$ axis. For $\phi$ = 0, corresponding to the backward-volume (BV) geometry, we make the standard assumption that the mode exhibits a uniform profile across the thickness. For~$\phi$ = $\frac{\pi}{2}$ (Damon Eshbash geometry) we assume that the MSSW mode profile can be described by an exponentially decaying function following \cite{damon_magnetostatic_1961-1} work. For intermediate angles, we postulate a mode profile as the combination of these two limits. Including an ellipticity factor calculated from Eq. (3.9) of \cite{verba_damping_2018}, the postulated mode profiles are written as:

\begin{equation}
      m_y^0=\epsilon(\cos \phi+\ e^{-kz}\sin\phi), ~    \epsilon m_z^0 = i m_y^0 ,~~~ \epsilon >1
      \label{postulat}
\end{equation}
where the exponent '0' is the mode index. This profile includes more physics, and would be expected to better describe the $\mu$-BLS spectra. This indeed improves the description but unfortunately not to a satisfactory level [see Fig.~\ref{Imag7}]. A likely reason is that for large-magnetization films with typical thicknesses between 10-60 nm, the MSSW mode does not have the exponential decaying profile \cite{kostylev_non-reciprocity_2013} of the exchange-free Damon-Eshbach approach, which it does for low-magnetic moment films such as YIG \cite{kostylev_non-reciprocity_2013}. 

\section{Conclusion}
In conclusion, our study on BiYIG, Co$_2$MnAl, and CoFeB illustrates how different magnetic materials exhibit distinct spectral features in their $\mu$-BLS spectra. The spectral shapes directly reflect the dispersive properties of the spin wave modes and therefore strongly depend on the film thickness. This work also discusses whether the use of numerical tools is necessary to compute the $\mu$-BLS spectra. Even if mode hybridization is not in the BLS wavevector sensitivity range, the Kalinikos and Slavin formalism may fail to accurately describe the spectra as illustrated by the CoFeB film, while it provides a good description for the BiYIG film.

\section*{Acknowledgement}
This work was supported by the French National Research Agency (ANR) as part of the “Investissements d’Avenir” and France 2030 programs. This includes the MARIN contract: ANR-20-CE24-0012 and the PEPR SPIN projects ANR 22 EXSP 0008 and ANR 22 EXSP 0004.

\bibliographystyle{IEEEtran}
\bibliography{IEEEabrv,refs,biblio2002}

\end{document}